\documentstyle[epsfig,eqsecnum,pre,multicol,aps]{revtex}
\newcommand{\bleq}{\ifpreprintsty
                   \else
                   \end{multicols}\widetext \vspace*{-3.5ex}{\tiny
                   
		\noindent\begin{tabular}[t]{c|}
                   \parbox{0.493\hsize}{~} \\ \hline \end{tabular}}
	                              \fi}
\newcommand{\eleq}{\ifpreprintsty
                   \else
                   {\tiny\hspace*{\fill}\begin{tabular}[t]{|c}\hline
                    \parbox{0.49\hsize}{~} \\
                    \end{tabular}}\vspace*{-2.5ex}\begin{multicols}{2}
	            \narrowtext
                    \fi}
\newcommand{\bcols}{\ifpreprintsty\else\begin{multicols}{2} 
	\narrowtext\fi}
\newcommand{\ecols}{\ifpreprintsty\else\end{multicols}\fi}
\begin{document}
\draft
\title{Nontrivial Polydispersity Exponents in
Aggregation Models}

\author{St\'ephane Cueille and Cl\'ement Sire}
\address{Laboratoire de Physique Quantique (UMR C5626 du CNRS),
Universit\'e Paul Sabatier\\
31062 Toulouse Cedex, France.\\}

\date{\today}
\maketitle
\begin{abstract}
We consider the scaling solutions of Smoluchowski's equation of irreversible
aggregation, for a non gelling collision kernel. The scaling mass distribution
$f(s)$ diverges as $s^{-\tau}$ when $s\to 0$.  $\tau$ is non trivial and could,
until  now, only be computed by numerical simulations. We develop here new {\em
general methods} to obtain exact bounds and  good approximations of $\tau$. For
the specific kernel $K_D^d(x,y)=(x^{1/D}+y^{1/D})^d$, describing  a mean-field
model of particles moving in $d$ dimensions and aggregating with conservation
of  ``mass'' $s=R^D$ ($R$ is the particle radius), perturbative and 
nonperturbative expansions are  derived.
 For a general kernel,   we find exact inequalities for $\tau$ and  
develop  a {\em variational approximation} which is used to carry out the 
first systematic study of  $\tau(d,D)$  for $K_D^d$.
The agreement is excellent both with the  expansions we derived and with 
 existing numerical values. 
Finally, we discuss a possible application to $2d$ decaying turbulence.
\end{abstract}


\bcols
\section*{Introduction}
Aggregation phenomena are widespread in Nature. They have such an impact on
material sciences, chemistry, astrophysics, that a large amount of literature
has been devoted to them \cite{friedlander,meakinrev,stanley,vicsek2}.
  In such dynamical processes,
particles or objects as  different in geometry and size as colloidal
particles, galaxies, small molecules, vortices in fluids, droplets, polymers,
 can merge
to form a new entity when they come into close contact or interpenetrate,
through diffusion (Brownian coagulation \cite{smolu,dlcca}), ballistic motion
(ballistic agglomeration \cite{pomeau,KRL,trizac}), exogenous growth (droplets
growth and coalescence \cite{meakindrop}) or droplet deposition
\cite{familymeakin}.

One is usually interested in the evolution of the statistical distribution of
the ``mass'' $s$, a quantity characteristic of each particle, that is conserved
in the  coalescence process: it can be either the actual mass, the volume, the
area, the electric charge, or any other physical quantity,
 depending on the underlying physics.

A great progress was achieved when it was proposed \cite{vicsek} and observed
both in real experiments and in numerical simulations that the mass 
distribution
$N(s,t)$ exhibits scale invariance at large time:
\begin{equation}
N(s,t)\sim S(t)^{-\beta} f\left(\frac{s}{S(t)}\right),\quad S(t)\sim t^z
\end{equation}
The divergence of the mass scale $S(t)$ bears on 
the oblivion of initial conditions and physical cut-off or
discreteness, as does the diverging correlation length of critical phenomena:
universality arises in dynamics as well, with new universality classes.

The exponents $z$ and $\beta$ are easily derived from conservation laws and
physical arguments, but in many cases a polydispersity exponent $\tau$ defined
by $f(x)\sim x^{-\tau}$ when $x\rightarrow 0$ is observed, whose value is
nontrivial though universal. The prediction of $\tau$ is still a challenge.

Except for a few (usually 1D) exactly solvable model \cite{takasire,privman},
 analytical results
are still lacking. The most popular, and the earliest, approach to these
 aggregation  problems is
Smoluchowski's equation \cite{smolu}, a master equation
\cite{vankam} for the one-body distribution $N(s,t)$:
\begin{eqnarray} \label{smoleq}
\frac{\partial N(s,t)}{\partial t}
&=&{\displaystyle \frac{1}{2} \int_0^s} 
N(s_1,t)N(s-s_1,t)K(s_1,s-s_1)\,ds_1\nonumber\\
&-& N(s,t) {\displaystyle \int_0^{+\infty}} N(s_1,t) K(s,s_1) \,ds_1
\end{eqnarray}
where the aggregation kernel $K(x,y)$ is symmetric and is characteristic of
the physics of the aggregation process on a more or less coarse-grained level.
Such kinetic equations are usually derived within a mean-field approximation,
where density fluctuations are ignored. Mean-field approximation
is expected to be valid above an upper critical spatial dimension.
 This dimension is usually $2$ for reaction-diffusion models, but van Dongen
showed that it can depend on the kernel \cite{vandong1}. Including some proper
approximation of the density-density correlations in the kernel
may improve Smoluchowski's approach \cite{poland}. 

Mean-field as it may be, Smoluchowski's equation is still highly nontrivial. 
No exact solution is available, except in a very few 
specific cases (see below),
and extracting  the nontrivial exponent $\tau$  for a
specific system from the proper kinetic equation is not an easy task.
 The problem was clarified by van Dongen and Ernst \cite{vandong2} who
classified the kernels according to their homogeneity and asymptotic behavior:
\begin{eqnarray}\label{genscal1}
K(bx,by)=b^\lambda K(x,y)\\
K(x,y)\sim x^\mu y^\nu \,\,\, (y \gg x) \label{genscal2}
\end{eqnarray}
For a given physical system, the homogeneity  $\lambda$ is easily determined
using scaling arguments. We consider only nongelling  systems with
$\lambda\leq1$ \cite{vandong2}. For $\mu>0$, the exponent $\tau$ is trivial
and found to  be $\tau=1+\lambda$, whereas for $\mu=0$, $\tau$ depends on the
whole solution $f$ of the scaling equation derived from Eq. (\ref{smoleq})
(see Eq. (\ref{scaleq}) below). $\mu<0$ does not lead to any  power law
behavior but rather to a bell-shaped scaling function $f$ \cite{vandong2}.

 In the following, we shall focus on the $\mu=0$ case  for which  the exponent
$\tau$ has so far only been determined numerically by direct simulation of
Smoluchowski's equation  \cite{kang,krivitsky},
 not an easy task \cite{meakinrev,kang}, by time series
\cite{song}, and of course by direct simulation of the physical system
supposed to be described by the considered Smoluchowski's equation 
\cite{meakinrev,dlcca,pomeau,trizac,meakindrop,familymeakin,vicsek,kang}. 
In the latter case, direct comparison with mean-field results is in principle
rather delicate. 
These methods are quite heavy, which explains that very few values of $\tau$ are
known \cite{krivitsky,song}, most of them concerning a specific kernel,
$K_D^d(x,y)=(x^{1/D}+y^{1/D})^d$ ($0\leq d \leq D$),  which appears in various
physical applications
\cite{meakinrev,stanley,vicsek2,poland,silk,ruckenstein,saffman}.

Considering the ubiquity and the importance of the $\mu=0$ case leading to
nontrivial polydispersity exponents, analytical results as well as more
effective numerical methods, making it possible to carry out extensive studies,
are certainly needed to use Smoluchowski's approach in a  predictive way. The
purpose of this article is to provide both and use 
them to perform the first complete study of $\tau(d,D)$ for the kernel
 $K_D^d=(x^{1/D}+y^{1/D})^d$. These analytical methods consist of exact 
bounds,  
perturbative and nonperturbative expansions around exactly solvable 
limits, while we introduce a {\em variational} scheme,
leading to excellent approximations of $\tau$ at extremely
 low computational cost,
without directly solving Smoluchowski's equation. We end the paper
with a practical application of our results in the field of 
two-dimensional turbulence.  

In section \ref{model}, we present a mean-field model of aggregation
of $D$-dimensional spheres diffusing in a $d$-dimensional space 
and coalescing with conservation of their volume, for which we derive a
Smoluchowski's equation with the kernel $K_D^d=(x^{1/D}+y^{1/D})^d$. Under 
the scaling hypothesis, we write down the equation for the scaling function,
 determine the exponents $z$ and $\beta$, and  derive an integral
equation for $\tau$ as well as a series of integral equations for the 
moments of the scaling function $f$. This section introduces hardly any 
new result and is intended merely to clarify notations, to present 
the state of the art, and to make a few useful
remarks.

Section \ref{sec.bounds} and \ref{sec.exp} present new analytical results
for the previously introduced kernel $K_D^d$. Section \ref{sec.bounds} 
describes 
a method to obtain exact bounds for any kernel, based on integral
equalities established in section \ref{model}.

  Section \ref{sec.exp} deals
with expansions of $\tau$ around its value for exactly solvable kernels.
Starting from the remark  that $K_D^d$ reduces to the constant kernel
 in both $d\to 0$
and $D\to \infty$ limits, for which an explicit exponential
solution is known,  we find some perturbative expansions in both limits. 
In the large $D$ limit with $d/D=\lambda$ fixed, the kernel reduces to 
$2^d(xy)^\lambda$ and we show that  $\tau \to 1+\lambda$, the first correction
being exponentially small  at large $d$, and thus nonperturbative.

In section \ref{sec.varia}, we present a variational approximation  based on
integral equations for the moments of $f$, and valid  for {\em any homogeneous
kernel}. This method reproduces some known exact results, and is used to compute
$\tau$ for a wide range of $d$ and $D$, the results being summarized on Fig.
\ref{fig.phase}. The approximation is compared to the few existing numerical
results \cite{krivitsky,song} as well as with   analytical expansions derived in
section \ref{sec.exp},  with excellent agreement and very low computational
cost.

Section \ref{sec.turb} presents a possible application in the field
of two-dimensional turbulence. We consider a model of 
diffusing and merging coherent vortices, and Smoluchowski's equation leads
to non Batchelor energy spectra with exponents in qualitative agreement with 
direct simulations found in the literature \cite{McWilliams,Benzi}.

\section{Model and scaling} \label{model}

Consider hyperspherical particles in a $d$-dimensional box, of polydisperse
radii $R$ with distribution $F(R,t)$, evolving the following way: at time $t$
we choose the positions of their centers with uniform probability in
$d$-space. Then each pair of overlapping spheres of radii $R_1$ and $R_2$
merges to form a new sphere of radius,
\begin{equation}\label{conser}
R=(R_1^{D}+R_2^{D})^\frac{1}{D}
\end{equation}
where $D$ is a parameter with $D\geq d$. $D$ can be the actual dimension of
the spheres, as for instance in the case of $D=3$ spheres deposited on a $d=2$
plane \cite{familymeakin}. Once each coalescence has been resolved, we have
reached time $t+\delta t$.

\subsection{Derivation of Smoluchowski's equation}

The conserved variable is $s=R^D$, and is continuous. We shall call $s_0$ the
physical lower cut-off, that is the charge of the smallest sphere in the initial
condition. Since the radius of a surviving sphere can only increase through
coalescence, $N(s,t)=0$ for $s<s_0$ and for any time $t>0$. Smoluchowski's
equation consists just in a balance of collisions. The number of collisions
between two spheres of radius $s_1^{1/D}$ and $s_2^{1/D}$ randomly and
independently deposited in the $d$ dimensional medium being
$N(s_1,t)N(s_2,t)\Omega_d(s_1^{1/D}+s_2^{1/D})^d$ where $\Omega_d$ is the
$d$-dimensional total solid angle. We obtain the equation,
\bleq
 \begin{eqnarray}
N(s,t+\delta t)-N(s,t) 
&\displaystyle=\Omega_d\left\{\frac{1}{2}\int_0^s N(s_1,t)
N(s-s_1,t)K_D^d(s_1,s-s_1)\,ds_1
\nonumber\right.\\
 &\displaystyle\left.- N(s,t) \int_0^{+\infty}
 N(s_1,t) K_D^d(s,s_1) \,ds_1\right\}
\end{eqnarray}
\eleq
with $K_D^d(x,y)=(x^{1/D}+y^{1/D})^d$.
We can get rid of the multiplicative constant, by properly choosing the time
unit $\delta t$ and by replacing the finite difference
 in time by a partial derivative to
exactly obtain Eq. (\ref{smoleq}).
 We notice that the
only approximation used to derive the equation is to neglect  multiple
collisions, for the system is intrinsically mean-field.
 
The kernel $K_D^d(x,y)=(x^{1/D}+y^{1/D})^d$ has been introduced 
in many contexts from molecular coagulation
\cite{poland} to cosmology \cite{krivitsky,silk} for specific values of $d$
and $D$, and is one of the most studied in the literature
\cite{poland,vandong2,krivitsky,song,silk,ziff,vandong3,Leyvraz} although very
few analytical results are known. This kernel has $\lambda=\frac{d}{D}$ and
$\mu=0$. Exact solutions are available in the case $d=0$ or $D=\infty$
 (constant kernel)
\cite{smolu}, and $d=D=1$ \cite{ziff}.

\subsection{Scaling} 
Now, we introduce the scaling form of $N(s,t)$. We first write the
conservation law. The total mass in the system is $\int_{s_0}^{+\infty}
sN(s,t) ds \sim  S(t)^{2-\beta} \int_0^{+\infty}  x f(x)dx$ and is conserved
which implies $\beta=2$, implicitly assuming that the
integral
of $xf(x)$ converges, i.e.,  in terms of the small $x$ divergence
 of $f$,  
that $\tau<2$, which will be shown below. 
We consider the total number of particles in the system $n(t)=\int_0^{+\infty}
N(s,t)\,ds$. It behaves at large time like
$S(t)^{1-\beta}\int_{s_0/S(t)}^{+\infty} f(x) dx$. If  $\,\tau<1$,  $n(t)\sim
S(t)^{1-\beta}\int_0^{+\infty}f(x)dx$ whereas if $\tau>1$, $n(t)\propto
S(t)^{\beta-\tau}$. If $\tau=1$, the integral diverges like $\ln (S(t))$,
hence $n(t)\propto S(t)^{1-\beta} \ln S(t)$. 

As promised, we are now able to show that $\tau<2$. if $\tau>2$, the total
charge in the system is proportional to $S(t)^{\tau -\beta}$, enforcing
$\beta=\tau$.  As a consequence, $n(t)$ would have a non zero  limit which is
impossible.   To summarize these results, we have, with $n(t)\propto t^{-z'}$,
\begin{eqnarray}
 \beta&=&2\\
  z'&=&\left\{ \begin{array}{lr}
	z,& \mbox{if }\tau<1\\
	z(2-\tau),& \mbox{if }\tau>1
       \end{array}\right.
\end{eqnarray}

The derivation of the scaling equation is rigorously described in
\cite{vandong4}, where it is shown that $S(t)\sim wt^z$, $w$ being some 
positive constant characteristic of the  time dependent equation. 
Plugging the scaling form of the distribution into Smoluchowski's
equation, and matching the large $t$  behavior of both sides of the
equation, yields $z=D/(d-D)$ and  the equation for the scaling function,
\begin{eqnarray}\label{scaleq}
&\displaystyle w\left[sf'(s)+2f(s)\right]
=f(s)\int_0^{+\infty} f(s_1)K_D^d(s_1,s) ds_1  \nonumber\\
&\displaystyle - \frac{1}{2}\int_0^s f(s_1)f(s-s_1)K_D^d(s_1,s-s_1) \,ds_1
\end{eqnarray}
If $\tau\geq 1$ each term of the RHS of Eq. (\ref{scaleq}) is separately
divergent and they should be properly grouped, for instance,
\bleq
\begin{eqnarray}
\displaystyle 
w\left[sf'(s)+2f(s)\right]&=&f(s)\int_{s/2}^{+\infty} f(s_1)K_D^d(s_1,s) ds_1  \nonumber\\
&-&\displaystyle \int_0^{s/2} f(s_1)\left[f(s-s_1)K_D^d(s_1,s-s_1)  - f(s)K_D^d(s_1,s)\right] \, ds_1,
\end{eqnarray}
\eleq
another way of taking care of these divergences  is to be found in 
\cite{vandong2,vandong4}.

As we are only interested in the exponent affecting  the small $s$ behavior of
$f$,  we shall set $w$ to unity  by changing $f$ to $wf$.  If $f(s)$ is a
solution of Eq. (\ref{scaleq}), then $b^{1+\lambda}f(b s)$ is also a solution.
The value of $b$ is often fixed by imposing   $\int\! xf(x)dx=1$,
 but we will
make a different choice for reasons that will become
clear later. 

A careful study of the large $s$ behavior of $f$ shows  that if $\lambda<1$ ($d<D$), 
$f(s)\approx c_\infty\, \delta \, s^{-\lambda} e^{-\delta s}$,
 with $c_\infty^{-1} =
\int_0^{1/2}K_D^d(x,1-x)
x^{-\lambda}(1-x)^{-\lambda}dx$ \cite{vandong4}. We choose 
the solution corresponding to $\delta=1$, which fixes $b$, and
leads to a nontrivial value for $\int\! xf(x)\,dx$. This asymptotic behavior
is not valid for $\lambda=1$ ($d=D$).  

For $d=0$ or $D=\infty$, Eq. (\ref{scaleq}) reduces to the constant kernel
equation with exact solution $f_0(x)=2e^{-s}$ and
$f_{\infty}(s)=2^{1-d}e^{-s}$ (note that the large $s$ asymptotics  become the
exact solution for all $s$ in these cases). 
For $d=1$ and $D=1$, an exact
analytic solution is also known for the time dependent equation,
the scaling function being $f(s)\propto s^{-3/2}e^{-s}$   \cite{ziff}, with
$z=\infty$ and $S(t)\propto e^{t}$.

Now, for given $d$ and $D$, and plugging the expected small $s$ behavior $f(s)
\sim s^{-\tau}$ into Eq. (\ref{scaleq}), one first gets that
$\tau<1+\lambda=1+d/D$. Then, matching the behavior of both sides of Eq.
(\ref{scaleq}) \cite{vandong2,vandong4}, one finds,
\begin{equation} \label{tau}
\tau=2-\int_0^{\infty} f(x) x^\lambda dx.
\end{equation}
If $\alpha>\tau-1$ we obtain by multiplying Eq. (\ref{scaleq}) by $x^\alpha$
 and integrating \cite{vandong2,vandong3},
\bleq
\begin{equation}\label{eqal}
 2(1-\alpha) \int_0^\infty x^\alpha f(x) \,dx=
 \int\!\!\int_0^\infty\!\!
 f(x)f(y)K_D^d(x,y)
\left[ x^\alpha+y^\alpha -\right.\left.(x+y)^\alpha\right]dx dy.
\end{equation}
\eleq
\subsection{Existing analytical and numerical results}\label{subsec.res}
Most existing analytical results for $\mu=0$ kernels are to be found in 
the beautiful series of papers by van Dongen and Ernst
\cite{vandong1,vandong2,vandong4,vandong3}.
 Apart from results mentioned earlier, they determined
the small $x$ subleading behavior of the scaling function, and they
found some inequalities for $\tau$ in the cases $d=1$, and $D=1$. 
In 1984, Leyvraz \cite{Leyvraz} proposed the analytical result $\tau=1+1/2D$
for the kernel $K_D^d$ with $d=1$, but in 1985, using exact inequalities,
van Dongen and Ernst showed 
that this result was erroneous and explained why it was so \cite{vandong3}. 
The argument of Leyvraz 
leading to this result is perfectly valid for class I kernels  with
$\mu>0$  for which
it predicts the correct exponent, but it breaks down for $\mu=0$ kernels.
We mention this fact for some references to the wrong result $\tau=1+1/2D$
can still  be found in some recent articles.   

We now review various kinds of numerical studies concerning the polydispersity
exponent $\tau$. These studies deal with the kernel $K_D^d$. 

Kang et al. \cite{kang} simulated a model of particle diffusion  and coalescence
(PCM) that can be shown to be exactly equivalent to  Smoluchowski's equation.
They also numerically directly computed the solution of the equation itself.
Their results concern the $d=1$ case. They   surprisingly found values of $\tau$
in contradiction with the exact bound $\tau \geq 1$ (see section
\ref{sec.bounds}) (for $D=4$, they found $\tau=0.63$). By comparison between
their two methods of computation, they concluded that in both cases they
observed a pseudo-asymptotic state, with wrong exponents but apparent scaling,
and that the actual asymptotic scaling regime appeared at  times too large to be
seen by their simulations. This illustrates the drawback of considering the
direct time evolution of the system: the actual asymptotic regime may not be
reached within the accessible  numerical simulation time scale. 
  
Krivitsky \cite{krivitsky} numerically solved Smoluchowski's equation
for the time dependent distribution for the kernel $K_D^d$, for $D=1,d\leq 1$,
 for which he determined 10 values for $\tau$ (see Fig. \ref{fig.smalld}).
Comparison with analytical results obtained by analysis of the scaling equation
(infinite time limit) in the present article will assess the fact that in
this case the asymptotic regime was actually reached by Krivitsky's solution.  
These numerical results will be found to be in excellent  agreement with our
variational method of section \ref{sec.varia}.

Song and Poland \cite{song}, computed  the large time evolution of the number of
clusters $n(t)\propto t^{-z'}$, and as $z'= z(2-\tau)$ when $\tau>1$, and
$z'=z$, when $\tau<1$, we can extract  $\tau$ from their data (for which
$\tau>1$). Their method consists in solving the equation  for $n(t)$ as a  power
series in time  $t$, and to extract the exponent $z'$ by manipulations of this
series. They treated only the cases $d=1,D=2$ and $d=2,D=3$. In the case
$d=1,D=2$, they present  two different results in the text (?). They first
consider $K_2^1$ and find $1/z'=0.57\pm 0.01$, then they extend their method two
$K_d^{d-1}$ and in the case $d=2$, which is exactly the same as previously, they
find $1/z'=0.588$ (they do not give  any error estimate in this case). In the
following, we shall see that we believe the first result to be  closer to the
exact one. In the next section, we shall see that their result in $d=2,D=3$
strongly violates exact inequalities, and thus is wrong.

The conclusion of this section, is that no complete study of the
value of $\tau$ had been performed until now because of a lack of 
appropriate numerical tools.  More precise analytical results would 
also certainly be welcome to guide numerical works. We see that simulating
or solving for the time evolution of the distribution function may not 
enable to reach the asymptotic scaling regime, and a guideline of the present
work will be to directly rely on the scaling equation corresponding
to the infinite time asymptotic state itself. 

\section{Exact bounds} \label{sec.bounds}
In the  next three sections, our workhorses will be both Eq. (\ref{tau}) and
(\ref{eqal}).

We first show that $\tau\geq 1$, for
$d\geq 1$. Suppose $\tau<1$ and consider Eq. (\ref{eqal}) with $\alpha=0$,
\begin{equation}\label{al=0}
2\int_0^{+\infty} f(x)dx= \int\!\!\int_0^{+\infty} f(x)f(y)K_D^d(x,y)\,dx\, dy.
\end{equation}
 For
$d\geq 1$, we have $(x^\frac{1}{D}+y^\frac{1}{D})^d\geq
x^\frac{d}{D}+y^\frac{d}{D}$, which leads to  $\int\! f(x)dx \geq \int\! f(x)dx
\int\! f(x)x^\frac{d}{D}dx$ (in the bulk of the text, all integrals should be
understood from 0 to $\infty$). Comparing with Eq. (\ref{tau}), this leads to
$1\geq 2-\tau$ or $\tau\geq 1$, which is contradictory. Notice that Eq.
(\ref{eqal}) with $\alpha=2$ for $d=1$ and $D=1$ leads to $\int\! x^2 f(x)dx= 2
(\int\! x^2f(x)dx)(\int\! xf(x)dx)$, and we recover the exact result
$\tau=2-\int\! x f(x)\,dx=3/2$ \cite{ziff} in a very simple way. These results
were already obtained by van Dongen and Ernst \cite{vandong3,vandong4}, who were
able to find  in the case $D=1$ the exact inequality,
$2d<\tau<2-2^{1-d}(1-d)/(2-2^d)$, which shows that $\tau=2d+O(d^2)$ when $d\to
0$. This interesting result will be generalized to any $D$ in next  section and
the $O(d^2)$ term will be computed in $D=1$. They  also found weaker
inequalities in $d=1$, but no result was obtained 
for general $d$ and $D$.

In order to deal with a general $\mu=0$ kernel $K$,
 we  introduce an extremely simple method 
to get lower and upper bounds
for $\tau$.  We rely on Eq. (\ref{eqal}) valid for $\alpha>\tau-1$. Combining
Eq. (\ref{tau}) and  (\ref{eqal}), we get:
\begin{equation} \label{mean}
\tau=2-(1-\alpha)\frac{\int\!\!\int_0^\infty g(x,y)\,dxdy}
{\int\!\!\int_0^\infty g(x,y)A(x/y)\,dxdy}
\end{equation}
where $A(u)=(1+u^\alpha-(1+u)^\alpha) K(x,y) /(u^\alpha+ u^\lambda)$
 satisfies $A(u)=A(1/u)$ and $g(x,y)=(x^\alpha y^\lambda+x^\lambda
y^\alpha)f(x)f(y)$. The ratio in Eq. (\ref{mean}) can then be interpreted as
the inverse of a kind of $average$ of $A(x/y)$ with the weight $g(x,y)$.  
For a given $\alpha \leq \lambda$, we numerically determine the maximum $M_\alpha$
and minimum $m_\alpha$ of the function $A(u)$.
Using Eq. (\ref{mean}),
this gives 
\begin{equation} \label{ineq}
2-(1-\alpha)/m_\alpha\leq \tau\leq 2-(1-\alpha)/M_\alpha
\end{equation}
 
 We then
choose the best values of $\alpha\leq \lambda$ compatible with $\alpha>\tau-1$
leading to the tightest bounds. More precisely, we proceed the following
way: we start with $\alpha=\lambda$ (as $\tau<1+\lambda$), 
from which we obtain some upper
and lower bound $\tau_{m}$ and $\tau_M$. If $\tau_M<1+\lambda$,
 Eq. (\ref{ineq})
holds for $\tau_M-1<\alpha\leq \lambda$, and we can compute new bounds
for each $\alpha$ in this interval, and find the tightest bounds.
 The upper bound obtained for $\alpha=\lambda$ cannot be improved
since $A(0)=1$ for $\alpha<\lambda$, hence $2-(1-\alpha)/M_\alpha \geq
1+\alpha$, but in many cases we can find  a better lower bound.

For $K_D^d$, a superficial plot of the function $A(u)$ may lead to the incorrect conclusion
that   its minimum is always obtained at $u=0$ with
$A(0)=1$. In fact a more careful study of $A$ shows that for certain values
of $\alpha$, the actual   minimum is at $u>0$ but very close to $0$.
For $u\to 0$, $A(u)\sim 1+du^{1/D}-u^{d/D-\alpha}$, and we see that if $\alpha
>(d-1)/D$, there is a local minimum for $u_m>0$ with $A(u_m)<1$.  For $d>1$,
and $\alpha=(d-1)/D+\varepsilon$, we get $u_m\sim \exp(-\ln(d)/\varepsilon)$, 
which vanishes
exponentially when $\varepsilon\to 0$ ($d>1$). Indeed, even when $\alpha$ is
not so close to $(d-1)/D$, $u_m$ may be very small. For instance, for 
$d=2,\,\,D=3$, and $\alpha=0.58598>(d-1)/D=0.333...$, we find that 
$u_m=1.365\times 10^{-4}$, and $A(u_m)=
0.7322$, which leads to a nontrivial lower bound of $1.4349$ for $\tau$.

Actually, it is easily seen that the inequalities obtained by van Dongen and
Ernst (in the case $d=1$ or $D=1$) correspond to $\alpha=d/D$ . In fact 
even in this case, $M_\alpha$ and $m_\alpha$ are nontrivial, and they used 
some {\em explicit} bounds of $M_\alpha$ and $m_\alpha$, which do not lead
to the tightest bounds for $\tau$. 

Thus, our method consists in computing the {\em actual} value  of 
$m_\alpha$ and $M_\alpha$, and 
varying $\alpha$ to optimize these bounds, 
which allows us to {\em greatly improve}  van Dongen and Ernst's explicit
inequalities  for $D=1$ or $d=1$, and to obtain new exact 
bounds for $d>1$. For instance, for the physically interesting cases (see
below) $(d=1,\,D=2)$, $(d=1,\,D=4)$ and $(d=2,\,D=4)$ we respectively found
$1.084\leq\tau\leq 1.147 $, $1\leq\tau\leq 1.075$ (compared to $1\leq\tau\leq
1.28 $ and $1\leq\tau\leq 1.109 $ in \cite{vandong3}) and $1.25\leq\tau\leq
1.5$. 

For $d=2,D=3$, we find $1.4349\leq\tau\leq 1.585$, which just  discards the
 value $\tau=1.244$ found by Song and Poland \cite{song}, and strongly
questions the validity of their approach. The exact bounds we obtained in $d=1,
D=2$ are violated by their alternative value $1.150$ for $\tau$ but not
by their first result $1.123$ (see subsection \ref{subsec.res}). 

It is useful to note that for any $D$, with $\alpha=d/D$, $A(u) \to 1/2$ when
$d\to 0$, which entails that $\tau \to 0$ (from Eq. (\ref{ineq})) in this limit.
 
To conclude with this topic of inequalities, 
let us consider 
Eq. (\ref{ineq}) with $\alpha=d/D$. In this case, when $D\to \infty$,
\begin{eqnarray}
A(u)&=&\frac{1}{2}(1+u^{-\frac{1}{D}})^d
\left[1+u^\frac{d}{D}-(1+u)^\frac{d}{D}\right]\nonumber \\
&\to& \left\{\begin{array}{lr}
		2^{d-1}, \quad 0<u\leq 1\\
	        \frac{1}{2}, \quad u=0
	\end{array}\right.	         
\end{eqnarray}
 hence  $m_\alpha\to 1/2$ and $M_\alpha \to 2^{d-1}$.
Therefore, the  upper   bound  for $\tau$
in  Eq. (\ref{ineq}) tends to $2-2^{1-d}$. This is strictly less than
 $1$ for $d<1$,
which means that for any $d<1$, there exists a 
finite critical $D_c(d)$, such that  $\tau<1$
 for any $D>D_c$. This result will be used in section \ref{sec.exp}.

\section{Perturbative  and nonperturbative expansions}\label{sec.exp}

 In this section  we use the exactly solvable
limits $d=0$ and $D=\infty$ as a basis for a perturbative expansion.
We also consider the case $d\to \infty$, keeping  $d/D=\lambda$ constant, for
which we find a nonperturbative expansion.

We saw that $\lim_{d\to 0}\tau=0$. What about the $D\to \infty$ 
limit of $\tau$ ? In fact, although strictly at $D=\infty$, 
$\tau$ is equal to $0$, as $f(x)=2^{1-d}e^{-x}$, we will see that 
$\tau_\infty=\lim_{D\to \infty}\tau>0$. This result was already noticed by 
van Dongen and Ernst in $d=1$ \cite{vandong3}. 
Since $\tau < 1+d/D$ we get that,
\begin{equation}
\tau_\infty\leq 1
\end{equation}
What can we learn from equation (\ref{tau}) in the large $D$ limit ?
We see that the limit for $\tau$ is
\begin{equation}
  \tau_\infty=2 - \int_0^{+\infty}\!\! f_\infty(x)\,dx= 2-2^{1-d} 
\end{equation}
provided that:
\begin{equation}
\label{integ1}
\lim_{D\to\infty} \int_0^{+\infty} (f_D(x)-f_\infty(x))\,x^\frac{d}{D}\, dx = 0
\end{equation}
For $d<1$, this result is consistent, since, from the last remark of section
\ref{sec.bounds}, we get $\tau_\infty\leq 2-2^{1-d}<1$.  

However, for $d \geq 1$   we know that $\tau \geq 1$, hence 
$\tau_\infty=1$, which means that for $d>1$,
\begin{equation}
\lim_{D\to\infty} \int_0^{+\infty} (f_D(x)-f_\infty(x)) \,x^\frac{d}{D} \,dx
 = 1-2^{1-d}> 0
\end{equation}  
while in $d=1$, (\ref{integ1}) is true.

Now that we know the large $D$ limit of $\tau$ ($\tau_\infty=1$ for $d>1$ and
$\tau_\infty=2-2^{1-d}$ for $d\leq 1$), as well as its small $d$ limit ($\tau\to
0$), let us compute the corresponding asymptotic corrections.
\subsection{Small $d$ expansion}
First, consider the limit $d \rightarrow 0$. We expand $f$ in series in
$d$: $f(x)=f_0(x)+d f_1(x) + O(d^2)$, $f_0(x)=e^{-x}$.
 A systematic way of expanding  $\tau$
would be to write down a linear (self-consistent) differential equation for
$f_1$ to solve it and plug the result into (\ref{tau}).

 However, as far as the first
order is concerned we can get it without solving for $f_1$. By developing the
integral expression of $\tau$, Eq. (\ref{tau}), we get,
\begin{eqnarray}
 \tau&=& 2-\int_0^{+\infty}\!\!
	f(x)x^\frac{d}{D} dx\nonumber\\
&=& -\frac{d}{D} \int_0^{+\infty}\!\! 
	f_0(x)\ln x \,dx - d\int_0^{+\infty}\!\! f_1(x)dx\nonumber\\
& & +O(d^2).
\end{eqnarray}
Then we 
expand both sides of Eq. (\ref{al=0})  to get an equation for
$\int\! f_1(x)dx$,
\begin{eqnarray}
\int_0^{+\infty}\!\! f_1(x)dx
&=& \frac{1}{2}\int\!\!\int_0^{+\infty}\!\! f_0(x)f_0(y)
\ln(x^\frac{1}{D}+y^\frac{1}{D})dx dy\nonumber \\
& &- \int_0^{+\infty}\!\! f_0(x)dx
 \int_0^{+\infty}\!\! f_1(x)dx
\end{eqnarray} 
 hence $\int\!\! f_1(x)dx=-\int\!\!\int e^{-x-y} \ln(x^\frac{1}{D}+y^\frac{1}{D}) 
dxdy$. After eliminating $\int f_1(x)dx$, we get:
\begin{eqnarray}\label{petitd}
\tau&=& 2d J_D  +O(d^2) \nonumber\\
J_D &=& \int_0^1 \ln\left(1+\left(\frac{1-u}{u}\right)^\frac{1}{D}\right)\,
du
\end{eqnarray}
Let us mention that this result can be systematically generalized to 
the case of any homogeneous kernel of
the form: $(g(x,y))^d$, leading to, $\tau= 2d\int_0^1 \ln g(1,\frac{1-u}{u})
du +O(d^2)$.

Although it may seem a  bit tedious, it is interesting to recover this result in
another way, as it shows that the small $x$ behavior of $f_1$ is consistent with
the $d\to 0$ expansion of the power law $x^{-\tau}=1-2dJ_D\ln x+O(d^2)$. Let us
write down the linear equation for $f_1$, 
\bleq
\begin{eqnarray}
xf_1'(x)+ 2 e^{-x}\int_0^x\!\!  f_1(y)e^{y}dy&=&2e^{-x} \int_0^{+\infty}\!\!
f_1(y)dy +4e^{-x}\int_0^{+\infty}\!\!e^{-y}\ln(y^{1/D}+x^{1/D})dy\nonumber\\
&-&2e^{-x} \int_0^{x}\!\!\ln(y^{1/D}+(x-y)^{1/D})dy.
\end{eqnarray}

With $u=e^xf_1$ we get the following equation:
\begin{eqnarray}\label{eqdifpert1}
x(u'-u)+2\int_0^x\!\!u(y)dy&=&2\int_0^{+\infty}\!\! u(y) e^{-y}dy+
4\int_0^{+\infty}\!\!e^{-y}\ln(y^{1/D}+x^{1/D})dy \nonumber \\
&-&2xJ_D-\frac{2}{D}(x\ln
x -x),
\end{eqnarray}
\eleq
which implies, after taking the derivative of Eq. (\ref{eqdifpert1}),
\begin{eqnarray}\label{eqdifpert2}
xu''+(1-x)u'+u&=&\frac{4}{D} \int_0^{+\infty}\!\!
e^{-y}
\frac{x^{1/D-1}}{y^{1/D}+x^{1/D}} dy \nonumber \\
& & -\frac{2}{D} \ln x-2J_D
\end{eqnarray}
the solution $u$ of (\ref{eqdifpert2}) involves two integration constants, one
being fixed by the fact that $f_1$ should go to zero at large $x$,
the other, $c_0$, by writing the compatibility with Eq. (\ref{eqdifpert1}), 
which can be done by taking the $x\to 0$ limit the latter equation.
From the expression of the solution (appendix \ref{append.sol}), or directly
from Eq. (\ref{eqdifpert2}), 
it is easily seen that $u$  has the asymptotic expansion for $x\rightarrow0$:
 \begin{equation}
 u(x)=b_0\ln x + O(1)
 \end{equation}
with $b_0=c_0-2/D$.

We know that $f(x)\sim cx^{-\tau}$ when $x\to 0$. When $d\to 0$, $c\to 2$ and
$\tau=d\tau_1+O(d^2)$, hence up to  order $d$ we expect, 
\begin{equation}
f(x)\sim 2\tau_1 \ln x 
\end{equation}
so that we interpret $b_0$ as $-2\tau_1$,  
\begin{equation}\label{taup}
\tau=-d\frac{b_0}{2}+O(d^2)
\end{equation}
The $x\to 0$ limit of (\ref{eqdifpert1}) is :
\begin{equation}\label{e1}
b_0 = 2\int_0^{+\infty}\!\! f_1(x)dx 
-\frac{4}{D}\int_0^{+\infty}\!\! e^{-x}\ln x \, dx  
\end{equation}
The integration of Eq. (\ref{eqdifpert2}) between $0$ and $+\infty$ yields,
\begin{equation}\label{e2}
-b_0+\int_0^{+\infty}\!\!f_1(x)dx= 2J_D -\frac{2}{D}\int_0^{+\infty}\!\!e^{-x}\ln x \, dx
\end{equation}
The combination of  Eqs. (\ref{e1}) and (\ref{e2}) yields $b_0$, which, substituted into Eq. (\ref{taup}), eventually leads to  the same result for $\tau$
 as previously obtained through the expansion of Eq. (\ref{al=0}) and 
Eq. (\ref{tau}).

For $D=1$, we get $\tau=2d+O(d^2)$, in good agreement with direct numerical
integration of Smoluchowski's equation performed by Krivitsky
\cite{krivitsky} and shown on Fig. \ref{fig.smalld} (see below).
 This result for $D=1$ also coincides up to order $O(d)$
with the  inequalities for $\tau$ that we obtained above, as noticed
in section \ref{sec.bounds}. This is not the case for other values of $D$.

The order $O(d^2)$ requires the computation of $f_1$.
However, in the special case $D=1$  it is possible
to obtain explicitly the $O(d^2)$ term by expanding Eq. (\ref{eqal}) for
$\alpha=d/D$ (see appendix \ref{d2}). We obtain,
\begin{equation}
\tau=2d+(\frac{\pi^2}{3}-4)d^2 +O(d^3)
\end{equation} 
In section \ref{sec.varia} (see Fig. \ref{fig.smalld}), we shall see
 that this result is in excellent agreement with both Krivitsky's results
and a new method of approximation that we shall introduce in section 
\ref{sec.varia}.

\subsection{Large D expansion} 
Now,  we perform an expansion in powers of $1/D$ for $d\leq 1$, expanding
$f(x)=f_{\infty}(x)+\frac{1}{D} f_1(x) + \frac{1}{D^2}f_2(x)+O(1/D^3)$. 

{\em Perturbative expansion in $d<1$} -
In $d<1$, as  mentioned in section \ref{sec.bounds},  $\tau<1$
 for any $D$ above a 
finite critical $D_c(d)$. As a consequence,  Eq. (\ref{eqal}) 
can be written for 
any $D>D_c(d)$. Therefore,  we can develop this equation for large $D$ in
powers of $1/D$, and  we find at first order, 
\bleq
\begin{equation}
\int_0^{+\infty}\!\! f_1(x)dx= d\,2^{d-2}\!\int\!\!\int_0^{+\infty}\!\!
 f_\infty(x) f_\infty(y)
(\ln x +\ln y)\,dx dy
+ 2^d\int_0^{+\infty}\!\! f_\infty(x)dx\int_0^{+\infty}\!\!
 f_1(x)dx, 
\end{equation} 
\eleq
hence :
\begin{eqnarray}\label{ord1}
\int_0^{+\infty}\!\! f_1(x)\,dx&=&-d\int_0^{+\infty}\!\! 
 f_\infty(x) \ln(x)\,dx \nonumber \\
&=& 2^{1-d}d\gamma,
\end{eqnarray}
where $\gamma$ is Euler's constant, while from Eq. (\ref{tau}),
\begin{eqnarray}
\tau&=& \tau_\infty -\frac{d}{D} \int_0^{+\infty}\!\! f_\infty(x) \ln(x) \,dx
 \nonumber \\ & &-\frac{1}{D} \int_0^{+\infty}\!\! f_1(x)dx+O(\frac{1}{D^2}).
\end{eqnarray}
We conclude, using Eq. (\ref{ord1}),  that the first order correction to
$\tau_\infty$ is zero.

The same method also gives access to the next term:
\begin{eqnarray}
\tau&=&\tau_\infty- \frac{d^2}{2D^2} \int_0^{+\infty}\!\!
 f_\infty(x) (\ln x)^2\,dx\nonumber \\
         & &-\frac{d}{D^2}\int_0^{+\infty}\!\!
 f_1(x)\ln(x)\,dx -\frac{1}{D^2} \int_0^{+\infty}\!\! f_2(x)dx,
\end{eqnarray}
while,
\bleq
\begin{eqnarray}
\int_0^{+\infty}\!\! f_2(x)dx&=& \frac{1}{2} \int\!\!\int_0^{+\infty}\!\! 
  f_\infty(x) f_\infty(y)
\frac{2^d}{8}\left[(d+1)((\ln x)^2 +(\ln y)^2)+2(d-1)\ln(x)\ln(y)\right]\,dxdy 
\nonumber\\
&+&2^{d-1}d\int\!\!\int_0^{+\infty}\!\!   f_1(x) f_1(y) (\ln x +\ln y)\,dx dy
\nonumber\\ &+&2^{d-1}{\left(\int_0^{+\infty}\!\! f_1(x)\, dx\right)}^2 
+2\int_0^{+\infty}\!\! f_2(x)\, dx
\end{eqnarray}
Using the known value of $\int f_1$ and our favourite integral table, we
get:
\begin{equation}
-\int_0^{+\infty}\!\! f_2(x)\, dx=
 \frac{d^2}{4} \int_0^{+\infty}\!\! f_\infty(x) (\ln x)^2 dx 
+d \int_0^{+\infty}\!\! f_1(x) \ln(x)
 dx + \frac {2^{1-d} d}{4}(\frac{\pi^2}{6}+d\gamma^2)
\end{equation}
\eleq
($\gamma$ being Euler's constant), which leads to:
\begin{equation} \label{alD2}
\tau =2-2^{1-d} + \frac{\pi^2 2^{-d} d(1-d)}{12D^2} +O(\frac{1}{D^3})
\end{equation}
Once again we were able to obtain a highly nontrivial expansion for $\tau$
without solving for $f_1$ and $f_2$ themselves, although this can also be
achieved this way. Note that in the limit of large $D$ $and$ small
$d$, Eq. (\ref{petitd}) and (\ref{alD2}) coincide up to order $O(d/D^2)$.

{\em Perturbative estimate for $d>1$ -} In the case $d\geq 1$, we have shown
that $\tau\geq 1$ and since $\tau<1+d/D$, we see that $\tau\to 1$ for
$D\to\infty$ and finite $d\geq 1$. As $f_1$ is non integrable, 
Eq. (\ref{eqal}) cannot be used with $\alpha=0$, and the previous perturbation
breaks down.

 Nevertheless we can try to obtain an estimate of
$\tau$ in the following way: we make the ansatz $f\sim
f_\infty+c/s^{1+\varepsilon} e^{-s}$. We plug it into
Eq. (\ref{tau}) and Eq. (\ref{eqal}) for $\alpha =d/D$, and after some algebra
(see appendix \ref{pestimate}) 
we see  that for consistency $\varepsilon$ must be of order $1/D$
and that $c=(1-2^{1-d})(d/D-\varepsilon)$, and eventually that
$\varepsilon=\kappa/D +O(1/D^2)$ where $\kappa$ is the solution of the
nonlinear equation:
\begin{equation}\label{estimate}
\frac{2}{1+2^{1-d}} = \int_0^1 (1+v^\frac{1}{d-\kappa})^d dv
\end{equation}

This equation always has a solution consistent with the exact bound
$1<\tau<1+d/D$. For instance in the case $d=2$, $D=4$ we obtain $\tau \approx
1.462$. Though it is still of order $1/D$, the obtained perturbative estimate
depends on the choice of $\alpha$. $\alpha=d/D$ seems however to be the most
natural choice.

In $d=1$, $c$ vanishes and we do not learn much. 
All terms of the $d<1$ series for $\tau$ in powers of $1/D$ vanish for
$d\rightarrow1$, as can be seen in Eq. (\ref{alD2}) for the two leading ones.
The reason is the following: the perturbation is derived from Eq. 
(\ref{al=0}) under the assumption that $\tau<1$. In $d=1$, such an assumption
 yields
$2\int f(x)dx= 2(\int x^{1/D}f(x) dx )(\int f(x)dx)$ hence $\tau=1$.
Consequently the
perturbative value of $\tau$ tends to $1$ when $d\to 1^-$. As will be
illustrated below by numerical results, for  a given $d>1$ there is 
a critical $D=D_c(d)$ above which $\tau<1$, and $D_c(d)$ tends to infinity 
when $d\to 1^-$, entailing the vanishing of the perturbation validity domain
in $D$. 
Thus, the correction to  $\tau=1$ for large $D$ may be {\it nonperturbative}
in $d=1$.

If we now take the $d\to\infty$ limit in Eq. (\ref{estimate}), we obtain,
$\tau\simeq 1+\lambda -2^{-d} \lambda$ ($\lambda =d/D$), a nonperturbative
behavior  in $d$ which is to be related to the  results below, obtained for
 $d\to \infty$,
$D\to\infty$, keeping  $\lambda$ constant.

\subsection{Large d and D}

We now present a nonperturbative calculation in the
limit of large $d$ $and$ $D$, keeping the ratio $\lambda=d/D$ fixed. In this
limit, the kernel can be written,
\begin{equation} \label{kerpro}
(x^\frac{1}{D} + y^\frac{1}{D})^d=2^d(xy)^{\frac{\lambda}{2}}(1+O(d/D^2))
\end{equation}
and surprisingly transforms into the well-studied ``product'' kernel
\cite{meakinrev,vandong2,vandong4,kang,krivitsky,song,Leyvraz}. 
Assuming scaling (a still
controversial subject \cite{krivitsky}), one can easily show that
$\tau=1+\lambda=1+d/D$ \cite{vandong2} (see also Eq. (\ref{genscal1}) and
(\ref{genscal2}) and the discussion below them, as it corresponds to
$\mu=\lambda/2>0$). 

We can show that including higher order corrections in
power of $1/D$ does not change the value of $\tau$ such that the correction to
$\tau=1+\lambda$ is certainly nonperturbative.
Consider the expansion of the kernel:
\begin{equation}
K(x,y)= 2^d(xy)^\frac{\lambda}{2}\left[ 1+ 2^{-d}O(1/d^2)\right]
\end{equation} 
The rescaled function $\tilde{f}=2^{d}f$ is the  solution
of the scaling form of Smoluchowski's equation with the kernel
$\tilde{K}=2^{-d}K(x,y)$, which is equal to $(xy)^\frac{\lambda}{2}$ 
{\em at every order in $1/d=1/(\lambda D)$}.
 In fact, we can estimate this
correction by assuming that for finite $d$ and $D$, 
\begin{equation}
\tilde{f}(s)\sim 
c_\lambda/s^{1+\lambda -\varepsilon_d}
\end{equation}
 for $s\to 0$. Plugging this estimate
into Eq. (\ref{tau}) with the limit kernel of Eq. (\ref{kerpro}), we first get
\begin{equation}
 \varepsilon_d\approx 2^{-d}\frac{c_\lambda}{(1-\lambda)}
\end{equation}
 $c_\lambda$ can be
determined by matching the coefficients of the leading terms in Eq.
(\ref{scaleq}) using the kernel of Eq. (\ref{kerpro}). After a straightforward
calculation, one gets  in the $d\to\infty$ limit,

\begin{eqnarray}
c_\lambda&=&{2(1-\lambda)}{I_\lambda}^{-1} \label{c}
\\
I_\lambda&=&\int_0^1
[u(1-u)]^{-1-\lambda/2}\left[u^\lambda+(1-u)^\lambda-1\right]\,du
\end{eqnarray}
which leads to 
\begin{equation} \label{nonpert}
\tau=1+\lambda-2^{1-d}I_\lambda^{-1}
\end{equation}
We thus find a nonperturbative (exponentially small) correction to $\tau$ in
the large $d$ and large $D$ limit, consistent with the result obtained above
for $d>1$ and large $D$. Note that Eq. (\ref{c}) is also consistent with the
exact result that $\tau\to 1$ as $D\to\infty$ for finite $d>1$, a result that
we obtain by setting $\lambda=0$ (as $I_\lambda$ diverges). 

\subsection{Summary of the results}
We have shown 
 that when $D\to \infty$,
 $\tau\to 1$ for $d\geq1$,
whereas $\tau\to 2-2^{1-d}<1$ for $d<1$. 
We were able to derive an $O(1/D^2)$ perturbative
expansion in $d<1$, and we convinced ourselves that the leading corrective
term in $d>1$ was of order $1/D$, by giving an estimate of this correction.
In $d=1$ both approaches break down and the large $D$ corrections to
 $\tau_\infty =1$ are probably nonperturbative. 

When $d\to 0$, $\tau$ goes to zero, and we gave a first order perturbative 
 expression in $d$, for any $D$. For $D=1$, we also found the explicit
coefficient in $d^2$. 

Eventually, we showed that for a fixed homogeneity $\lambda=d/D$, 
$\tau$ tends exponentially to $1+\lambda$ at large $d$. In the following
section we present a new general numerical method to compute $\tau$ and
we confirm our analytical result by performing the first extensive study
of the function $\tau(d,D)$.

\section{Variational approach}\label{sec.varia}
In this section, we present a practical way of obtaining good approximate 
values for $\tau$, without explicitly solving  Smoluchowski's equation. 
  Once again, we rely on Eq. (\ref{eqal}), which holds for the exact 
scaling function (solution of Eq. (\ref{scaleq})),
 for any $\alpha>\tau-1$. This equation
is {\em general}, and does not depend on the specific kernel we
 study in this article.
 As a consequence, the methods we develop are general and do apply to
 {\em any homogeneous kernel}. We emphasize the fact that this method
does not intend to approach the whole scaling function, but sets the focus on
the computation of $\tau$ (in fact, numerically solving the scaling 
equation Eq. (\ref{scaleq}) for the scaling function seems to be  at least 
as  difficult as directly solving the time-dependent equation \cite{moi}). 

\subsection{Principles of the method}
The simplest way of approximating $\tau$ is to  evaluate the ``average'' in
Eq. (\ref{mean}) using a reasonable trial weight function $g(x,y)$ instead of
the unknown exact one. As a simple start, we will expose a crude, 
but straightforward algorithm, that illustrates the basic idea.
 Then we will develop the variational method itself, which is not much more
intricate, but much more effective.

A one parameter  choice for a trial weight function is obtained
 by replacing in the
above expression of $g(x,y)$ the exact $f(x)$ by $f_\tau(x)=x^{-\tau}\exp(-x)$
 which has
the correct leading asymptotics for small $x$ (by definition of $\tau$)
 and decays exponentially at large $x$, although not with the exact asymptotics 
$x^{-\lambda}e^{-x}$ ($\lambda<1$) \cite{vandong1}. Still, this functional 
form is known
to be a good approximation of the actual $f(x)$ obtained in simulations
\cite{krivitsky},  and is even the exact solution, but  for a 
multiplicative constant, for the constant kernel ($\tau=0$) and in the
  $d=D=1$ case,
which belongs to the special class $\lambda=1$ \cite{ziff}. The first idea
that comes to mind
is just to determine $\tau$ self-consistently such that Eq. (\ref{mean})
holds for $f_\tau$, with a specific choice of $\alpha$, 
for instance   $\alpha=\lambda$. 
This is readily done, by an iterative method: starting from an initial $\tau_0$, verifying previously obtained exact bounds, we construct the sequence.
\begin{equation}
\tau_{n+1}=(1-\varepsilon)+\varepsilon\left(2-(1-\alpha)R_\alpha(f_{\tau_n})
\right)
\end{equation} 
with \begin{equation}
      R_\alpha(\phi)= 2\frac{\int\!\!\int_0^{+\infty} x^{\alpha}\phi(x)
y^\lambda
 \phi(y)dx dy}{\int \int_0^{+\infty} \phi(x)\phi(y) K(x,y)
\left[(x+y)^\alpha-x^\alpha-y^\alpha\right]}
     \end{equation}
which converges, with a proper choice of $1>\varepsilon>0$, to 
a fixed point corresponding to an $f_\tau$ verifying Eq. (\ref{mean}).
The numerical evaluation of $R(\tau)$ can be achieved with utter celerity
and arbitrary precision, since it reduces to the calculation of 
one-dimensional integrals, and of a few values of the gamma function, thanks
to a very convenient transformation (see
 appendix \ref{use}). We notice that it is unnecessary to include any 
  multiplicative constant into $f_\tau$, since it would just cancel out
in Eq. (\ref{mean}).

Of course,  this algorithm should yield different values
 of $\tau$ for different
choices of $\alpha$,
 except in the special case when the exact solution is of the form
$f_\tau$. This corresponds to $d=0$, $D=\infty$ and $d=1,D=1$, and
this method converges by construction, to the exact value of $\tau$,
 but for the round-off errors.  
In the generic case, the variation can be non negligible
(in $d=2,D=4$, $\tau\approx 1.371$ for $\alpha=d/D$, while
$\tau\approx 1.398$ for $\alpha=0.403$)) and 
  the fixed point $\tau$ may even violate exact bounds.
For instance, in the case $d=1,D=3$ with $\alpha=d/D$ we get $\tau=0.9894$
whereas we know that $\tau>1$. The variation with $\alpha$ makes the
method unreliable. In $d=2,D=4$, it gives $\tau\approx 1.385\pm 0.015$,
compared to $\tau\approx 1.434 \pm 0.004$  
with  the variational approximation, that we now introduce,
 which, starting from
the same basic idea, proves to be much more effective.

{\em Variational approximation -}
A much better and hardly more intricate method is to choose a reasonable sample
of values of $\alpha$, and minimize an error function measuring the violation of
the corresponding Eq. (\ref{mean}). This method can be systematically improved
by allowing for $n$ free `fitting' parameters (including $\tau$ itself) in the
trial weight $g(x,y)$. In the following we will proceed by replacing the  exact
$f$ by a variational function of the form,
\begin{equation}
 f_v(x,\tau_0,\tau_1,..\tau_n,c_1..c_i)=x^{-\tau_0}e^{-x}+ \sum_{j=1}^{n}
c_jx^{-\tau_j}e^{-x}
\end{equation}
and we will minimize the error function, 
\begin{equation}
 \chi^2(f_v)= \sum_i\left(\tau_0-2+(1-\alpha_i)R_{\alpha_i}(f_v)\right)^2   
\end{equation}
to get a variational approximation $\tau_v=\tau_0$ of $\tau$.
Brute force should not be used in the evaluation of $\chi^2$: once again,
 Eq. (\ref{utile}) makes it possible to drastically reduce
the computation time, and to perform the evaluation of $\chi^2$ with 
an excellent precision.

Of course, the values of the exponents in $f_v$ should not be blindly chosen.
van Dongen and Ernst \cite{vandong4} showed that the subleading term
in the small $x$ asymptotic expansion of $f$ is
\begin{equation}
 \propto \left\{\begin{array}{ll}
	 x^{1+\lambda-2\tau}, & \mbox{if } \tau>1+\lambda- \mu_1,\\
         x^{\mu_1-\tau}, & \mbox{if } \tau<1+\lambda- \mu_1\\
	 x^{-\tau}\ln x, & \mbox{if } \tau=1+\lambda- \mu_1\
       \end{array}
\right.	
\end{equation}	
with $K(x,y)-x^\lambda \propto y^{\mu_1}x^{\lambda-\mu_1}$ when $x\to \infty$,
whereas the exact asymptotic at large $x$  is $\propto x^{-\lambda} e^{-x}$.
Therefore, a good three-parameters class of trial functions should be:
\begin{equation}
f_v(x,\tau_0,c_1,c_2)=\left(\frac{1}{x^{\tau_0}}
+ \frac{c_1}{x^{\tau_1(\tau_0)}}
+ \frac{c_2}{x^\lambda}\right)
e^{-x}
\end{equation}
$\tau_1$ being either $2\tau_0-1-\lambda$ (if $\tau_0>1+\lambda-\mu_1$), or
$\tau_0-\mu_1$ (if $\tau_0<1+\lambda-\mu_1$).
 The small $x$ leading term in $f_v$
is $\tau_0$ provided that $\tau_0>\lambda$.
The approximate value $\tau_v$ is the value of $\tau_0$ at the minimum.

By construction, this method reproduces the exact results  for the constant
kernel and $d=1,D=1$, since the exact scaling function  is contained in those
cases in the class of variational function we chose. In general, this method is
inadequate to approach $f$ itself,  and is just designed to compute $\tau$, in
the same way as the  variational approach in quantum mechanics is designed to
obtain eigenvalues  but, in principle, not eigenfunctions.
\subsection{Implementation} With a small number $n$ of variational parameters,
we choose to perform  the minimization with the downhill simplex method
described 
in \cite{numrec}
 (steepest descent, conjugate gradient or other methods could
also be used, with the drawback  that these methods require extra evaluations
of $\chi^2$ to compute its gradient). This method starts from
a $n$-dimensional simplex, i.e. $n+1$ points in the $n$-dimensional 
parameter space, and performs a sequence of geometric deformations until
it contracts to a local minimum of the function. It is not the fastest
algorithm, but it easily converges, and in our case where the computational
 burden is low we do not need more sophisticated devices.  

As in any optimization problem, the initial condition is a crucial 
parameter, but here there is the additional complication 
that the smallest moment $\alpha_{min}$ used in the computation
of $\chi^2$, should be bigger than $\tau-1$, and bigger than $\tau_0-1$
at any step of the algorithm. What information on the value of $\tau$
we may  a priori gather  
(exact bounds, perturbation expansion), should guide our choice. 
Anyway, we do know that
$\tau<1+\lambda$: starting with an initial $\tau_0$ smaller than 
$1+\lambda$ and $\alpha_{min}>\lambda$ should avoid any trouble.
As we get a first approximation of $\tau$ we will be able to decrease
the value of $\alpha_{min}$ and make it closer to $\tau_v-1$,
while refining the initial conditions. A few Monte-Carlo
minimization steps can also be used to find a proper initial condition 
(but we scarcely needed this functionality in this work).

Why should we choose as small an $\alpha_{min}$ as possible ? 
The answer is that small moments probe the small $x$ divergence of
$f(x)$, which is precisely what we are interested in. However, we
also need some intermediate and higher moments to probe the 
intermediate $x$ and the large $x$ decay to stabilize consistent
values of $c_1$ and $c_2$. There should be at least as many moments 
as variational parameter, otherwise there would be an infinite number
 of minima. 
Too many moments would cause excessive numerical round-off errors in the 
computation of $\chi^2$.

We tested round-off errors by computing $\tau_v$ for the exactly
solvable model $K_1^1$ for which $f(x)\propto x^{-3/2} e^{-x}$, since, 
were we endowed with infinite numerical precision, our algorithm would
yield the exact result in this case, as said before, whatever the 
$\alpha_i$ may be, provided that they all are bigger than $1/2=\tau-1$.

With the three parameter function introduced above, and moments
$0.55$, $0.667$, $0.783$, $0.9$ and $2$, we find $\tau=1.49997\pm 4\times 
10^{-6}$ ($\chi^2=1.94\times 10^{-8}$), the uncertainty being due to 
variations with different choices for the initial values of the parameters
 and the tolerance on the
size of the simplex (the minimization algorithm stop criterion).
The round-off errors increase with the number of moments and
the number of variational parameters. The error is much bigger on 
$c_1$ and $c_2$, we find $c_1=0.11 \pm 0.1$ and $c_2=-0.12 \pm 0.1$,
instead of strictly $0$. This means that the sensitivity on $c_1$ and $c_2$ is
small in the vicinity of the minimum, and this method is not the 
right one to determine the scaling function 
(a negative $c_2$ is unphysical here), but  it just was not devised for this
purpose: we just meant to compute $\tau$, and for this quantity the accuracy
 is excellent.

\subsection{Numerical results}
We used this method to determine approximations of $\tau$ for
the kernel $(x^{1/D}+y^{1/D})^d$. We compared our results
to numerical values obtained for $d\leq 1$, $D=1$ by Krivitsky 
\cite{krivitsky}, and to our perturbative and nonperturbative
expansions.

All values were obtained from the three-parameter variational functions
introduced earlier  in this text. We used 8 moments, 6 in the interval
$[\alpha_{min},0.9]$, plus $\alpha=2$ and $\alpha=3$. $\alpha_{min}$ was
adjusted to be as close to $\tau_v-1$ as possible.
 The computation time was from
$1$ to $10$ seconds per run on a HP workstation. $2$ to $5$ runs per points
were necessary to adjust the parameters.

We also computed a few points with a different repartition of moments:
5 in the range $[\tau-1,d/D]$, $\alpha=0.9$, $2$, $3$, as well as 
with only $2$ variational parameters ($c_1=0$), and with $4$ variational
parameters (the additional exponent being $\mu_1-\tau$ in the case when
$\tau>1+(d-1)/D$). The observed {\em relative} variations
 of $\tau_v$ were at most of a few $10^{-3}$. In all cases, $\tau$ was
found to be consistent with exact bounds.

\begin{figure}
\begin{center}
\epsfig{figure=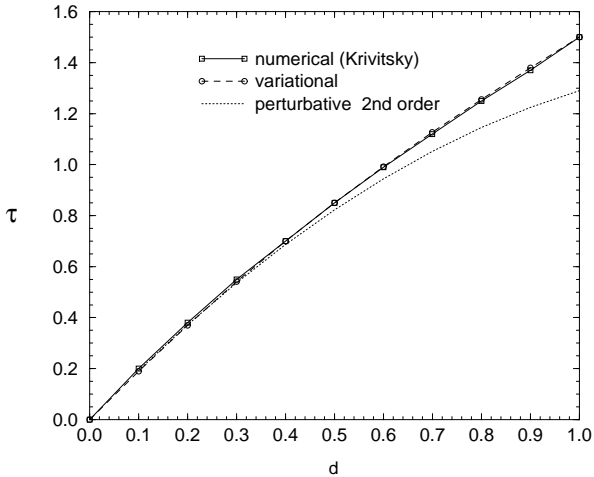,width=\linewidth}
\caption{In $D=1$, the comparison between the results obtained in
 {\protect \cite{krivitsky}} by Krivitsky,
the variational approximation with 3 parameters and 8 moments, and the $O(d^2)$
perturbative expansion of $\tau$, illustrates the efficiency of the variational
approximation. Indeed, the agreement between the numerical solution of
Smoluchowski's equation {\protect \cite{krivitsky}}  and the variational
approximation is excellent. The variational approximation is even in closer
agreement with the small $d$  perturbative expansion than Krivitsky's result,
and although both methods recover the exact result $\tau=3/2$ for $D=1$,
Krivitsky's curve seems 
to have an accident in the vicinity of $D=1$, whereas the variational
 result is smooth.}
\label{fig.smalld}
\end{center}
\end{figure}

First, we consider the case $D=1$. Fig. (\ref{fig.smalld}) shows the
 comparison
between variational approximations of $\tau$ obtained 
with the modus operandi
we just exposed, values extracted by Krivitsky \cite{krivitsky}
 from a numerical solution
of Smoluchowski's equation, and the $O(d^2)$ perturbative expansion. 
The agreement between the variational approximation and Krivitsky's
results is excellent, which confirms the effectiveness and  
efficiency of the method:
the ratio computation time (a few seconds)/accuracy is impressive.
Actually, the variational approximation looks smoother than Krivitsky's
curve, which has two visible accidents (small cusps) near $d=1$ and $d=0.4$,
and the variational approximation is fully consistent with the exact 
$O(d^2)$ expansion at small $d$ to which it clearly tends asymptotically,
whereas Krivitsky's result tends to remain parallel to the perturbative curve,
though close to it. Its good agreement with our infinite time results 
assesses the fact that Krivitsky's solution actually
 reached the scaling regime,
which, as said in section \ref{model}, was not obvious a priori. 
We conclude that in this regime, the variational approximation recovers and
confirms  the results obtained by numerical integration of Smoluchowski's
equation.
   
Once the effectiveness of the method was established, we were able to
 carry out the first systematic study of $\tau(d,D)$, and to control its
validity thanks to the analytical results obtained in sections \ref{sec.bounds}
and \ref{sec.exp}. 

\begin{figure}
\begin{center}
\epsfig{figure=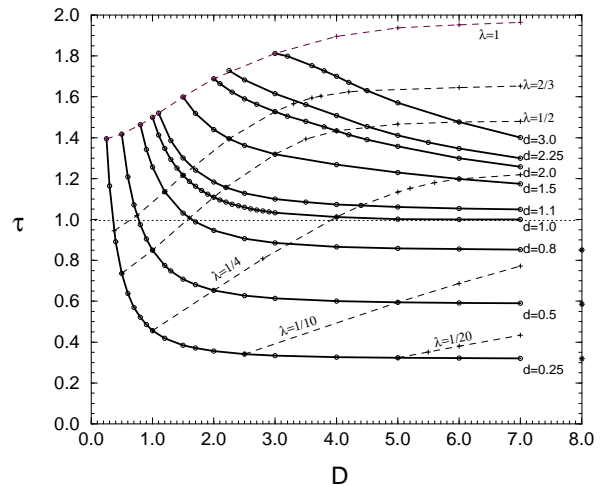,width=\linewidth}
\caption{The exponent $\tau$ was computed by the variational method for various
values of $d$ and $D$. We show here some iso-$d$(solid lines) and iso-$\lambda$
(dashed) ($\lambda=d/D$)
lines. The iso-$d$ lines tend to $\tau=2-2^{1-d}$(stars on the right axis) 
if $d<1$, and  to $1$ 
if $d\geq 1$. The critical $D$ above which $\tau$ becomes smaller than $1$ 
tends to infinity when $d\to 1^{-}$, entailing the breakdown of the large
$D$ perturbative expansion in $D\geq 1$. The $d=1$ iso-$d$ line seems to
tend exponentially to $1$, while for $d>1$ the relaxation to $1$ is slower.
An inflexion point appears above $d\approx 2$. 
The iso-$\lambda$ lines exponentially saturate to $1+\lambda$ at large $D$.}
\label{fig.phase}
\end{center}
\end{figure}

We show on  Fig. \ref{fig.phase}, the function $\tau(d,D)$ ($0.25\leq d\leq 3$,
$d\leq D<8.$) plotted
in a $(\tau,D)$ diagram. Two kinds of curves are shown.
Solid lines represent some iso-$d$ lines, i.e the function $\tau(D)$ for
a fixed value of $d$, whereas  dashed lines are  iso-$\lambda$ ($\lambda=d/D$)
lines. The reliability of the approximation is assessed by the comparison
with analytical results. As established in section \ref{sec.exp}
iso-$d$ lines tend to $\tau=2-2^{1-d}$ (stars on the right axis of 
Fig. \ref{fig.phase}) 
if $d<1$, and  to $1$ 
if $d\geq 1$. As expected, the critical $D$ above which 
$\tau$ becomes smaller than $1$ 
tends to infinity when $d\to 1^{-}$, entailing the breakdown of the large
$D$ perturbative expansion in $D\geq 1$. The $d=1$ iso-$d$ line seems to
tend exponentially to $1$, which is consistent with a nonperturbative  
decay in $1/D$ (see below).
 For $d>1$ the large $D$ decay is slower as analytically
predicted (we found a $1/D$ perturbative correction, see below).
For $d\geq2$  the curves shape qualitatively changes and an inflexion point 
appears. 

\begin{figure}
\begin{center}
\epsfig{figure=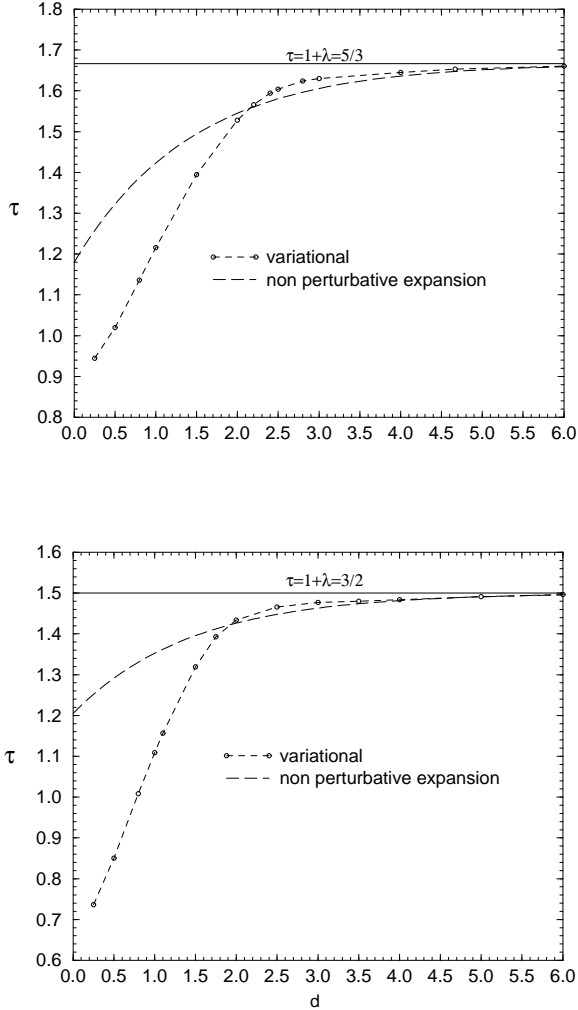,width=\linewidth}
\caption{Iso-$\lambda$ curves computed by the variational method 
(solid lines), as a function of $d$, for $\lambda=1/2$ 
and $\lambda=2/3$. As analytically established, $\tau$ tends to 
$1+\lambda$ at large
$D$. The agreement is good at large $d$ with the nonperturbative 
expansion (dashed lines).}
\label{fig.isolambda}
\end{center}
\end{figure}

 Iso-$\lambda$ lines exponentially saturate to $1+\lambda$ at large $D$,
as analytically established before.
 Fig. \ref{fig.isolambda} shows the comparison 
between the variational approximation and the nonperturbative large $d$
expansion of Eq. (\ref{nonpert})
in two cases, $\lambda=1/2$ and $\lambda=2/3$. The agreement is once again 
excellent at large $d$.   

\begin{figure}
\begin{center}
\epsfig{figure=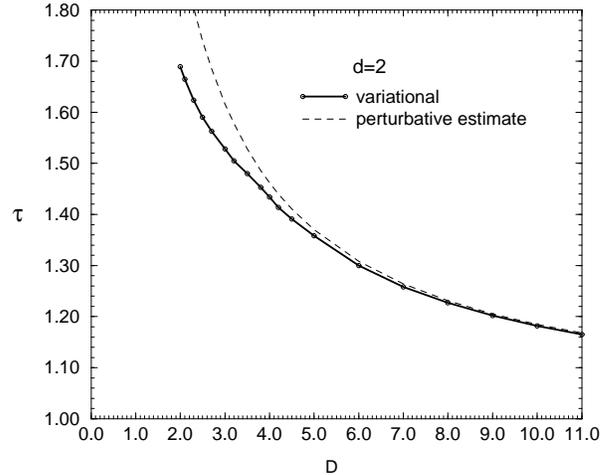,width=\linewidth}
\caption{In $d=2$, the exponents computed by the variational
approximation are in good agreement with the perturbative large $D$ 
estimate $\tau=1+1.849/D$. From data, the actual asymptotic correction
seems to be closer to $1.82/D$. The cusp on the variational curve corresponds
to the change of behavior with the occurrence of an inflexion point 
for above $d=2$.}
\label{fig.pert}
\end{center}
\end{figure}

\begin{figure}
\begin{center}
\epsfig{figure=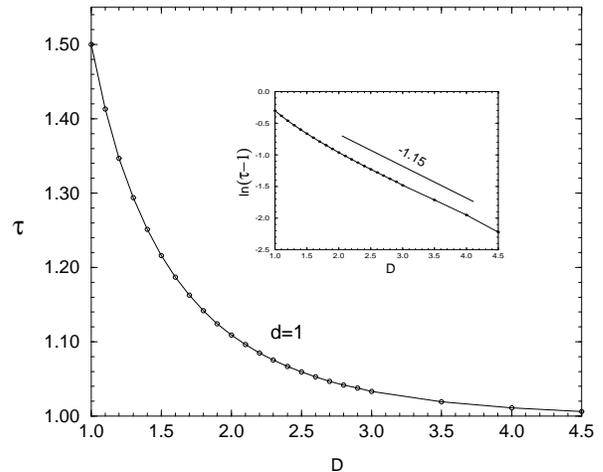,width=\linewidth}
\caption{For $d=1$, the exponents computed by the variational
approximation displays a much faster decay to their $D=\infty$ limit 
($\tau_\infty=1$), than for $d>1$. 
Indeed, as shown on this figure, the decay seems 
to be exponential in $D$, with roughly $\tau-1\propto e^{-1.15 D}$,
a nonperturbative behavior 
to be related to the break-down of the large $D$ 
perturbative approaches for $d=1$.}
\label{fig.d1}
\end{center}
\end{figure}

In $d=1,D=2$, Song and Poland \cite{song} found $\tau=1.123\pm 0.016$ (using
their first result), which compares well with our $\tau=1.109$. In $d=2,D=3$,
we find $\tau=1.528$ which, unlike their result ($1.243$),
 is perfectly consistent
with the exact bounds $1.4349<\tau<1.585$. In $d=2,D=4$, we find $\tau=1.434$,
which is in fair agreement with the perturbative large $D$ estimate 
$\tau=1.462$ of section \ref{sec.exp}. In fact, as shown on Fig.
\ref{fig.pert}, the perturbative estimate is indeed a good approximation
of $\tau$ in $d=2$ for $D\leq 6$, and the $\propto 1/D$ decay is confirmed
by the variational results. The cusp on the variational curve 
is confirmed by the existence of an inflexion point on $d>2$ curves, as 
mentioned above. In $d=1$, a nonperturbative exponential large $D$ decay to
$\tau_\infty=1$, is confirmed by Fig. \ref{fig.d1}. We roughly 
find $\tau-1 \propto e^{-1.15 D}$.  

\begin{figure}
\begin{center}
\epsfig{figure=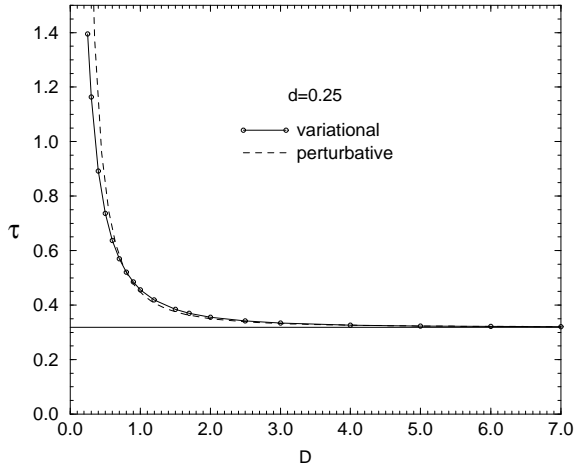,width=\linewidth}
\caption{In $d=0.25$, the exponents computed by the variational
approximation are in good agreement with the perturbative large $D$ 
estimate $\tau=2-2^{1-d}+ \frac{\pi^2 2^{-d}d(1-d)}{12D^2}+O(1/D^3)$.}
\label{fig.D2}
\end{center}
\end{figure}

Eventually, we show on Fig. \ref{fig.D2} (for $d=0.25$), that
the variational result is also in good agreement with the large $D$ 
second order perturbative expansion in $d<1$ ($\propto 1/D^2$).

As this section draws to a close, we shall say that this variational method,
 although very simple,
 seems to be very well adapted to the determination of the exponent 
$\tau$, as it is fast 
and, at least in the case we studied in this article, very accurate.
It made it possible to acquire for the first time  quantitative
 knowledge of $\tau$ 
in the whole parameter space of the $K_D^d$ kernel, the most studied and 
the prototype of the notorious class II kernels. The method is general
and could help shedding some  light on the whole class of kernels, thus
 increasing the practical use of Smoluchowski's approach to understand 
aggregation phenomena. This point is worth an example. This is 
precisely what is dealt with  in  section \ref{sec.turb}. 
\section{Application in two-dimensional decaying turbulence}
\label{sec.turb}
In this section, we would like to illustrate the results obtained 
in this article by presenting an
original application outside the field of massive particle aggregation,
namely the dynamics of vortices in two-dimensional decaying turbulence.

Recently, a statistical numerical model was introduced \cite{McWilliams,Benzi}
which describes the dynamics and the merger of vortices with the assumption
that the typical core vorticity $\omega$ and the total energy $E\sim \int
v^2\,d^2x\sim \sum_i\omega^2 R_i^4$ are conserved ($R_i$ is the radius of the
$i$-th vortex) throughout the merging processes. This model reproduces the
main features observed in direct numerical simulations (see
\cite{McWilliams,Benzi} for details). For instance, after noting that a
distribution of vortex radii satisfying $P(R)\sim R^{-\beta}$ is equivalent to
a Gaussian energy spectrum $E(k)\sim k^{\beta-6}$ \cite{Benzi}, the simulation
of this model was able to reproduce the fact that starting from a Batchelor
spectrum $E(k)\sim k^{-3}$ ($\beta=3$), the system evolves systematically to a
steeper spectrum $E(k)\sim k^{-\gamma}$ with $\gamma=6-\beta$ in the range
$\gamma\approx 3\sim 5$ \cite{Benzi}.

Now, one expects that the collision kernel between two vortices is somewhat
intermediate between the ballistic hard-disk form $\sigma\sim (R_1+R_2)$
\cite{song}, and the totally uncorrelated form $\sigma\sim (R_1+R_2)^2$ (where
the probability of colliding is proportional to the probability that two
randomly placed vortices overlap, see also below Eq. (\ref{conser})).
 Thus, one can describe approximately the decay of vortices due to
mergers by means of Eq. (\ref{scaleq}) with $1\leq d \leq 2$ and $D=4$, as two
colliding vortices merge into a new one with $R=(R_1^{4}+R_2^{4})^{1/4}$ in
order to conserve energy and core vorticity. One thus expects a power law
radius distribution $P(R)\sim R^{-\beta}$, with $\beta=D(\tau-1)+1$ and $\tau$
given by our model.  We find values of $\gamma$ ranging from $\gamma\approx
3.26$ for $d=2$ (taking $\tau=1.434$) to $\gamma\approx 4.95$ (taking
$\tau= 1.012$) for $d=1$, in good qualitative agreement with observed
exponents. As also found in direct simulations, the actual exponent (and here
the value of the effective correct $d$) could depend on the actual initial
conditions ($\omega$, area occupied by the vortices $\sim$ enstrophy). Note
that the Batchelor limit case $\gamma=3$, is obtained when taking the naive
strict upper bound $\tau=1+d/D$ with $d=2$ and $D=4$.

\section*{Conclusion}
In this article, we tackled the notoriously difficult problem of 
nontrivial polydispersity exponents in Smoluchowski's approach to 
aggregation from an original angle. We chose to directly start 
from the scaling (infinite time limit) equation, 
and we did not focus on the determination of
the whole scaling function, which is the object of solving Smoluchowski's
equation, to concentrate on $\tau$ itself, which actually mainly depends
on global  (integral) equations. We think, and illustrated this point
on the example of a simplified model of two-dimensional turbulence, that 
in some cases, the only knowledge of $\tau$ would still be a good step
towards the understanding of the phenomenon. The choices we made were fruitful
and  gave birth to new analytical and numerical  results. 

From an analytical viewpoint, we were able to use integral equations 
to find some exact bounds for $\tau$, and, in the specific case of $K_D^d=
(x^{1/D}+y^{1/D})^d$, we obtained some perturbative and nonperturbative
expansions of $\tau$, without explicitly computing the corresponding expansions
for the whole scaling function.

From a numerical viewpoint, we devised a variational approximation scheme, that
recovers by construction known exact results, and can be used as a tool for
extensive determination of $\tau$, since it is both very  economical and
accurate. In addition, it is likely that the scaling function obtained in the
variational approach is in many cases qualitatively, if not quantitatively,
right.  To illustrate its effectiveness, we performed a comprehensive study of
$\tau$ for a wide range  of the parameters $(d,D)$ of the kernel $K_D^d$. This
is  a noticeable advance, since 
 very little quantitative knowledge was available for this kernel,
although it was the prototype  kernels 
with a nontrivial $\tau$, and  the object of much
 attention in the past 
\cite{poland,vandong2,kang,krivitsky,song,silk,ruckenstein,saffman,ziff,vandong3,Leyvraz}.

\acknowledgements
We are very grateful to Jane Lion Basson, Fran\c{c}ois Leyvraz and Sid Redner
for valuable comments.

\appendix
\section{A useful formula}\label{use}
\bleq
\begin{eqnarray}\label{utile}
\int\!\!\int_0^{+\infty} x^{-\tau_1} y^{-\tau_2} e^{-x-y} K(x,y) 
\left[(x+y)^\alpha-x^\alpha-y^\alpha\right]dx dy=\nonumber\\
\Gamma(2+\lambda+\alpha-\tau_1-
\tau_2) \left[{\cal X}(\tau_1,\alpha,\tau_1+\tau_2) +{\cal X}(\tau_2,\alpha,
\tau_1+\tau_2)\right],
\end{eqnarray}
\eleq
where $\Gamma$ is the gamma function, and,
\begin{equation}
{\cal X}(t,\alpha,q)=\int_0^1 \frac{K(1,u)\left[(1+u)^\alpha-1-u^\alpha\right]}
{u^t(1+u)^{2+\lambda+\alpha-q}}du
\end{equation}

To demonstrate this formula is straightforward: just make the
 change of variable
$x=uv, y=v$, and use the definition of the $\Gamma$ function:
\begin{equation}
\Gamma(x)=\int_0^{+\infty} t^{x-1}e^{-t}dt
\end{equation}

From a numerical viewpoint this formula makes it possible to implement 
very rapid and accurate code for the variational approximations we developed
before. It would be very awkward and inefficient to use 2-dimensional 
numerical integration (especially here, as the integrand is singular at the 
origin). A startlingly economical way of computing the gamma function
is due to Lanczos and is described in \cite{numrec} (it is not much 
slower than the built-in exponential function...). 

\vspace{0.5cm}
\section{The $O(d^2)$ term in D=1}\label{d2}

We derive the $O(d^2)$ correction to $\tau=2d$ for $D=1$, by computing the $d^2$
order of  respectively Eqs. (\ref{tau}) and (\ref{eqal}) with $\alpha=d$, to
get,
\bleq
\begin{eqnarray}
4-2a_2&=& c+ 4\int_0^{+\infty}\!\! f_1(x)\ln x dx 
+4 \int_0^{+\infty}\!\! f_2(x)dx \label{a1}\\
 -a_2 &=& \frac{1}{2} \int_0^{+\infty}\!\! e^{-x}(\ln x)^2dx 
+\int_0^{+\infty}\!\! f_1(x)\ln x dx + \int_0^{+\infty}\!\! f_2(x) dx,
\label{a2}
\end{eqnarray} 
where $\tau=2d+a_2 d^2+O(d^3)$, and,
\begin{eqnarray}
c&=&4 \int_0^{+\infty}\!\! e^{-x} (\ln(x))^2 dx +
4\int\!\!\int_0^{+\infty}\!\! e^{-x-y}\ln(x+y)\ln\frac{xy}{x+y} dx dy 
\nonumber\\
 &+& 4 \left( \int_0^{+\infty}\!\! f_1(x)\,dx \right)
\left(\int_0^{+\infty}\!\! e^{-y}\ln y \,dy\right)
+\left(\int_0^{+\infty}\!\! f_1(x)\, dx\right)^2.
\end{eqnarray}
\eleq
$c$ can be computed since $\int f_1$ is known from the first order calculation.
After some elementary transformations, we find that $c-4\int e^{-x}
 (\ln x)^2dx =\frac{2\pi^2}{3}-4$. Combining Eq. (\ref{a1}) and 
(\ref{a2})), we find $4+2a_2=c-4\int e^{-x} (\ln x)^2 dx$, hence eventually
\begin{equation}
a_2=\frac{\pi^2}{3}-4.
\end{equation}

\section{The linearized scaling function} \label{append.sol}
We find the solution of the second order differential equation 
Eq. (\ref{eqdifpert2}) for the linear coefficient $f_1(x)$ in the small
 $d$ expansion
of the scaling function. With $u(x)=e^{x}f_1(x)$, the latter equation is  
\begin{eqnarray}
xu''+(1-x)u'+u&=&\frac{4}{D} \int_0^{+\infty}\!\!
e^{-y}
\frac{x^{1/D-1}}{y^{1/D}+x^{1/D}} dy\nonumber\\
& & -\frac{2}{D} \ln x-2J_D.
\end{eqnarray}
With $v(x)=u(x)/(x-1)$, this equation reduces to a first order differential
 equation  for $v'$, and we find,
\bleq
\begin{eqnarray}
f_1(x)&=& c_0\, u_0(x)e^{-x}+ c_1(x-1)- 2J_D-\frac{2}{D}(1+\ln x)\nonumber\\
&+& \frac{4}{D}e^{-x}\int_0^{x}\!\!dy_1 \frac{e^{y_1}}{y_1(y_1-1)^2}\int_0^{y_1}\!\! dy_2\,\,
y_2^{1/D-1}e^{-y_2}(y_2-1)\int_0^{+\infty}\!\!dy_3 \frac{e^{-y_3}}{y_3^{1/D}+
y_2^{1/D}},
\end{eqnarray}
\eleq
and 
\begin{equation}
  u_0(x)=  e^x-(x-1) \mbox{Vp}\left(\int_{-\infty}^x \frac{e^y}{y} dy \right)
\end{equation}
(``Vp'' means  ``principal value'').

In fact, the triple integral can be transformed into  a simple integral
involving special functions. For our purpose, we only need to know that
this integral goes to zero when $x\to 0$, which is easily seen.

\section{Perturbative estimate} \label{pestimate}
For $d>1$, $\tau=1+\varepsilon(D)$ where $\varepsilon\to0$ when $D\to\infty$.
We make the ansatz:
\begin{equation}
f(x)\approx f_\infty(x) + \frac{c}{x^{1+\varepsilon}} e^{-x},
\end{equation}
and plug it into Eq. (\ref{tau}) to obtain, $1-\varepsilon= 2^{1-d}+ 
c\,\Gamma(\frac{d}{D}-\varepsilon)$, which means that, when $D\to\infty$,
$c\approx (1-2^{1-d})(d/D-\varepsilon)$. 
Then we make use of Eqs. (\ref{utile}) and (\ref{eqal}) to obtain,
\bleq
\begin{eqnarray}
  2(1-\frac{d}{D})(1-\varepsilon)&=& 2^{2-2d} \int\!\!\int e^{-x-y}
  (x^\frac{1}{D}
+y^\frac{1}{D})^d [x^\frac{d}{D}+y^\frac{d}{D}-(x+y)^\frac{d}{D}]dx dy\nonumber
\\
&+&2^{2-d} c \quad \Gamma(1+2d/D-\varepsilon)\left[
{\cal X}(0,d/D,1+\varepsilon)+ {\cal X}(1+\varepsilon,d/D,1+\varepsilon)\right]
\nonumber\\
&+& 2c^2\Gamma(2d/D-2\varepsilon) {\cal X}(1+\varepsilon,d/D, 2+2\varepsilon)
\end{eqnarray} .
\eleq
	
The next step is to write down the limit of this equation when
$D\to\infty$. We know that $\Gamma(x)\sim_{x\to 0} 1/x$, and a change of
variable $v=u^{d/D-\varepsilon}$ in the integral factors ${\cal X}$ shows that
${\cal X}(1+\varepsilon,d/D,1+\epsilon)\sim{\cal
X}(1+\varepsilon,d/D,2+\varepsilon)\sim (d/D-\varepsilon)^{-1}
\int_0^1(1+v^\frac{1}{d-D\varepsilon})^ddv$. We obtain:
\begin{eqnarray}
 2&=&2^{2-d} +2^{2-d}
 \frac{c}{d/D-\varepsilon}\int_0^1(1+v^\frac{1}{d-\kappa})^ddv\nonumber \\
& & + \frac{c^2}{(d/D-\varepsilon)^2} \int_0^1(1+v^\frac{1}{d-\kappa})^ddv
\end{eqnarray}
$\kappa$ is the limit of $D\varepsilon$. Taking into account the value of
$c$, we finally get,
\begin{eqnarray}
\frac{2}{1+2^{1-d}}=\int_0^1 (1+v^\frac{1}{d-\kappa})^ddv=J(\kappa,d) \label
{kappa}\\
\tau=1+\frac{\kappa}{D}+O(\frac{1}{D^2})
\end{eqnarray}
The equation Eq. (\ref{kappa}) has a unique solution $0<\kappa<d$ since
the integral $J(\kappa,d)$ is a decreasing function of $\kappa$, and
$J(0,d)=2^d>2/(2^{1-d}+1)>1=J(d,d)$  (for $d>1$).

\ecols

\end{document}